\documentclass[aps,prl,twocolumn,showpacs,superscriptaddress,groupedaddress]{revtex4}

\usepackage{amsthm}
\usepackage{graphicx}
\usepackage{amsfonts}
\usepackage{amssymb}
\usepackage{times,txfonts}
\usepackage{hyperref}
\usepackage{color}
\newcommand{\ket}[1]{|#1\rangle}
\newcommand{\bra}[1]{\langle#1|}
\newcommand{\BD}{${\rm{M}}^3_2$ }
\newcommand{\NBD}{${\rm{M}}^3_N$ }

\begin{document}

\title{Frozen Quantum Coherence}

\author{Thomas R. Bromley}
\email{pmxtrb@nottingham.ac.uk}
\affiliation{$\mbox{School of Mathematical Sciences, The University of Nottingham, University Park, Nottingham NG7 2RD, United Kingdom}$}

\author{Marco Cianciaruso}
\email{cianciaruso.marco@gmail.com}
\affiliation{$\mbox{School of Mathematical Sciences, The University of Nottingham, University Park, Nottingham NG7 2RD, United Kingdom}$}
\affiliation{Dipartimento di Fisica ``E. R. Caianiello'', Universit\`a degli Studi di Salerno, Via Giovanni Paolo II, I-84084 Fisciano (SA), Italy; and INFN Sezione di Napoli, Gruppo Collegato di Salerno, Italy}

\author{Gerardo Adesso}
\email{gerardo.adesso@nottingham.ac.uk}
\affiliation{$\mbox{School of Mathematical Sciences, The University of Nottingham, University Park, Nottingham NG7 2RD, United Kingdom}$}

\begin{abstract}
We analyse under which dynamical conditions the coherence of an open quantum system is totally unaffected by noise. For a single qubit, specific measures of coherence are found to freeze under different  conditions, with no general agreement between them. Conversely, for an $N$-qubit system with even $N$, we identify universal conditions in terms of initial states and local incoherent channels such that all bona fide distance-based coherence monotones are left invariant during the entire evolution. This finding  also provides an insightful physical interpretation for the freezing phenomenon of quantum correlations beyond entanglement.  We further obtain analytical results for distance-based measures of coherence in two-qubit states with maximally mixed marginals.
\end{abstract}

\date{May 27, 2015}
\maketitle







\noindent\emph{\bfseries Introduction.} The coherent superposition of states stands as one of the characteristic features that mark the departure of quantum mechanics from the classical realm, if not  {\it the} most essential one \cite{Leggett1980}.  Quantum coherence constitutes a powerful resource for quantum metrology \cite{Giovannetti2004,MacconeNew} and entanglement creation \cite{Asboth05,StreltsovSingh}, and is at the root of a number of intriguing phenomena of wide-ranging impact in quantum optics \cite{Glauber1963, Scully1991,Albrecht1994,BWallsMilburn}, quantum information \cite{Nielsen2010}, solid state physics \cite{BookSolid,Li2012}, and thermodynamics \cite{Ford1978,Correa2014,Rossnagel2014,Narasimhachar2014,Lostaglio2014, Aaberg2014}. In recent years, research on the presence and functional role of quantum coherence in biological systems has also attracted a considerable interest
\cite{Engel2007, Plenio2008, Aspuru2008,Aspuru2009,Aspuru2009b, Caruso2009, Panitchayangkoon2010, Collini2010, Alexandra2011, Lloyd2011, Chin2012,Lambert2013, Huelga2013, Chin2013, Cai2013, Alexandra2014, Giorgi2015}.


Despite the fundamental importance of quantum coherence, only very recently have relevant first steps been achieved towards developing a rigorous theory of coherence as a physical resource \cite{Baumgratz2014,Levi2014,Marvian2013}, and necessary  constraints have been put forward to assess valid quantifiers of coherence \cite{Baumgratz2014,Aaberg2006}.  A number of coherence measures have been proposed and investigated, such as the $l_{1}$-norm and relative entropy of coherence \cite{Baumgratz2014}, and the skew information \cite{Girolami2014,Spekkens2014}. Attempts to quantify coherence via a distance-based approach, which has been fruitfully adopted for entanglement and other correlations \cite{Vedral1997, Plenio2007,  Horodecki2009,Modi2010, Modi2012,Spehner2013,Nakano2013,Sarandy2013,Aaronson2013, Bromley2014, Cianciaruso2014}, have revealed some subtleties \cite{Shao2014}.


\begin{figure}[t]
  \centering
    \includegraphics[width=8.5cm]{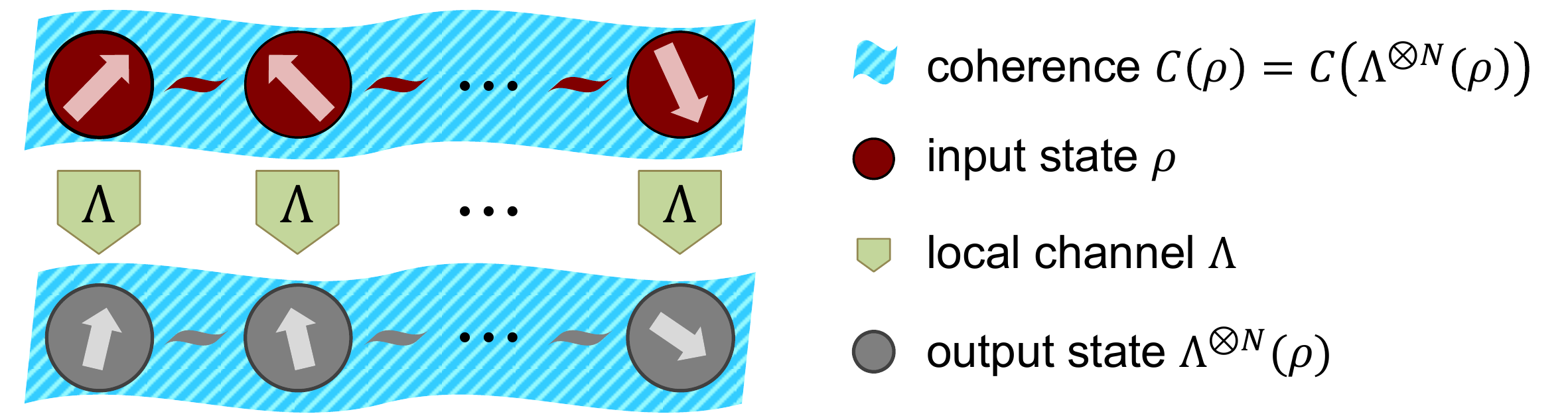}
 \caption{
(color online) Frozen quantum coherence for an $N$-qubit system subject to incoherent noisy channels $\Lambda$ acting on each qubit.}
 \label{frolla}
\end{figure}


A lesson learned from natural sciences is that coherence-based effects can flourish and persist at significant timescales under suitable exposure to decohering environments. Recent evidence suggests that a fruitful interplay between long-lived quantum coherence and tailored noise may be in fact crucial to enhance certain biological processes, such as light harvesting \cite{Alexandra2011,Lloyd2011,Lambert2013,Huelga2013}. This surprising cooperation between traditionally competing phenomena provides an inspiration to explore other physical contexts, such as quantum information science, in order to seek for general conditions under which coherence can be sustained in the presence of typical sources of noise \cite{Walls1985,Monras2014}. Progress on this fundamental question can lead to a more efficient exploitation of coherence to empower the performance of real-world quantum technologies.

In this Letter we investigate the dynamics of quantum coherence in open quantum  systems under paradigmatic incoherent noisy channels. While coherence is generally nonincreasing under any incoherent channel \cite{Baumgratz2014}, our goal is to identify initial states and dynamical conditions, here labelled {\it freezing conditions}, such that coherence will remain exactly constant (frozen) during the whole evolution (see Fig.~\ref{frolla}).

For a single qubit subject to a Markovian bit flip, bit-phase flip, phase flip, depolarising, amplitude damping, or phase damping channel \cite{Nielsen2010}, we study the evolution of the $l_{1}$-norm and relative entropy of coherence \cite{Baumgratz2014} with respect to the computational basis. We show that no nontrivial condition exists such that both measures are simultaneously frozen. We then turn our attention to two-qubit systems, for which we remarkably identify a set of initial states such that {\it all} bona fide distance-based measures of coherence are frozen forever when each qubit is independently experiencing a nondissipative flip channel. These results are extended to $N$-qubit systems with any even $N$, for which suitable conditions supporting the freezing of all distance-based measures of coherence are provided. Such a {\it universal} freezing of quantum coherence within the geometric approach is intimately related to the freezing of distance-based quantum correlations beyond entanglement \cite{Maziero2009,Mazzola2010, Aaronson2013, Cianciaruso2014, Chanda2014}, thus shedding light on the latter  from a physical perspective. Finally, some analytical results  for the  $l_{1}$-norm of coherence are obtained, and its freezing conditions in general one- and two-qubit states are identified.




\smallskip
\noindent\emph{\bfseries Incoherent states and channels.}
Quantum coherence is conventionally associated with the capability of a quantum state to exhibit quantum interference phenomena \cite{BWallsMilburn}. Coherence effects are usually ascribed to the off-diagonal elements of a density matrix with respect to a particular reference basis, whose choice is dictated by the physical scenario under consideration \cite{notebasis}.
Here, for an $N$-qubit system associated to a Hilbert space $\mathbb{C}^{2^N}$, we fix the computational basis $\{\ket{0},\,\ket{1}\}^{\otimes N}$ as the reference basis, and we define incoherent states as those whose density matrix $\delta$ is diagonal in such a basis,
\begin{equation}\label{incoherent}
\delta = {\sum}_{i_1, \ldots, i_N=0}^1 d_{i_1, \ldots, i_N} \ket{i_1,\ldots,i_N}\bra{i_1,\ldots,i_N}\,.
\end{equation}

Markovian dynamics of an open quantum system is described by a completely positive trace-preserving (CPTP) map $\Lambda$, i.e.~a quantum channel, whose action on the state $\rho$ of the system can be characterised by a set of Kraus operators $\{K_j\}$ such that $ \Lambda (\rho) = \sum_{j} K_{j} \rho K_{j}^{\dagger}$,
where $\sum_{j} K_{j}^{\dagger} K_{j} = \mathbb{I}$. Incoherent quantum channels (ICPTP maps) constitute a subset of quantum channels that satisfy the additional constraint  $K_{j} \mathcal{I} K_{j}^\dagger \subset \mathcal{I}$ for all $j$, where $\mathcal{I}$ is the set of incoherent states \cite{Baumgratz2014}.
This implies that ICPTP maps transform incoherent states into incoherent states, and no creation of coherence would be witnessed even if an observer had access to individual outcomes.

We will consider paradigmatic instances of incoherent channels which embody typical noise sources in quantum information processing \cite{Nielsen2010, Baumgratz2014}, and whose action on a single qubit is described as follows, in terms of a parameter $q\in[0,1]$ which encodes the strength of the noise. The bit flip, bit-phase flip and phase flip channels are represented in Kraus form by
\begin{equation}\label{Eq:FlipChannels}
K^{F_k}_0 = \sqrt{1-q/2}\ \mathbb{I},\,
K^{F_k}_{i,j\neq k}=0, \,
K^{F_k}_k = \sqrt{q/2}\ \sigma_{k},\,
\end{equation}
with $k=1$, $k=2$ and $k=3$, respectively, and $\sigma_j$ being the $j$-th Pauli matrix. The depolarising channel is represented by
$K^D_{0} = \sqrt{1-3q/4}\mathbb{I},\,
K^D_{j} = \sqrt{q/4} \sigma_{j},$
with $j \in \{1,2,3\}$. Finally, the amplitude damping channel is represented by
$K^A_{0}=\left( \begin{array}{cc}
1 & 0 \\
0 & \sqrt{1-q} \end{array} \right),\,
K^A_{1}=\left( \begin{array}{cc}
0 & \sqrt{q} \\
0 & 0 \end{array} \right),\,$
and the phase damping channel by
$K^P_{0}=\left( \begin{array}{cc}
1 & 0 \\
0 & \sqrt{1-q} \end{array} \right),\,
K^P_{1}=\left( \begin{array}{cc}
0 & 0 \\
0 & \sqrt{q} \end{array} \right).$

The action of $N$ independent and identical local noisy channels (of a given type, say labelled by $\Xi=\{F_k, D, A, P\}$) on each qubit of an $N$-qubit system, as depicted in Fig.~\ref{frolla}, maps the system state $\rho$ into the evolved state
\begin{equation}\label{eq:independentchannels}
\Lambda_q^{\Xi \otimes N}(\rho) = \sum_{j_1,\cdots,j_N} \big(K^\Xi_{j_1}\otimes\cdots\otimes K^\Xi_{j_N}\big)\ \rho\ \big({K^\Xi_{j_1}}^\dagger\otimes \cdots\otimes {K^\Xi_{j_N}}^\dagger\big)\,.
\end{equation}

\smallskip
\noindent\emph{\bfseries Coherence monotones.}
Baumgratz \emph{et al.} \cite{Baumgratz2014} have formulated a set of physical requirements which should be satisfied by any valid measure of quantum coherence ${\cal C}$, namely: \[
\mbox{
\begin{tabular}{p{.075\columnwidth} p{.77\columnwidth}}
    {\it C1.} & ${\cal C}(\rho)\geq 0$ for all states $\rho$, with  ${\cal C}(\delta)=0$ for all incoherent states $\delta \in \mathcal{I}$; \\
    {\it C2a.} & Contractivity under incoherent channels  $\Lambda_{\rm ICPTP}$, $ {\cal C}(\rho) \geq {\cal C}(\Lambda_{\rm ICPTP} (\rho))$; \\
    {\it C2b.} & Contractivity under selective measurements on average, ${\cal C}(\rho)\geq \sum_{j} p_{j} {\cal C}(\rho_{j})$, where $\rho_{j}=K_{j} \rho K_{j}^{\dagger} / p_{j}$ and $p_{j}=\text{Tr}(K_{j} \rho K_{j}^{\dagger})$, for any $\{K_{j}\}$ such that $\sum_{j}K_{j}^{\dagger} K_{j} = \mathbb{I}$ and $K_{j} \mathcal{I} K_{j} \subset \mathcal{I}$ for all $j$; \\
    {\it C3.} & Convexity, ${\cal C}(q \rho + (1-q)\tau) \leq q {\cal C}(\rho)+(1-q) {\cal C}(\tau)$ for any states $\rho$ and $\tau$ and $q \in [0,1]$.
\end{tabular}
}
\]
We now recall known measures of coherence.  The $l_{1}$-norm quantifies coherence
in an intuitive way, via the off-diagonal
elements of a density matrix $\rho$ in the reference basis \cite{Baumgratz2014},
\begin{equation}\label{eq:l1def}
{\cal C}_{l_{1}}(\rho) = {\sum}_{i \neq j} \left| \rho_{ij} \right|.
\end{equation}

Alternatively, one can quantify coherence by means of a geometric approach. Given a distance $D$, a generic distance-based measure of coherence is defined as
\begin{equation}\label{eq:CDdef}
{\cal C}_{D} (\rho) = \min _{\delta \in \mathcal{I}} D (\rho, \delta) = D (\rho, \delta_{\rho})\,,
\end{equation}
where  $\delta_{\rho}$ is one of the closest incoherent states to $\rho$ with respect to $D$. We refer to bona fide distances $D$ as those which satisfy natural properties \cite{Nielsen2010} of contractivity under quantum channels, i.e.~$D(\Lambda (\rho),\Lambda (\tau)) \leq D(\rho,\tau)$ for any states $\rho, \tau$ and CPTP map $\Lambda$, and joint convexity, i.e.~$D(q \rho+(1-q)\varpi, q \tau + (1-q)\varsigma)\leq q D(\rho,\tau) + (1-q)D(\varpi,\varsigma)$
for any states $\rho, \varpi, \tau, \varsigma$ and $q \in [0,1]$. We then refer to bona fide distance-based measures of coherence ${\cal C}_D$ as those defined by Eq.~(\ref{eq:CDdef}) using a bona fide distance $D$: all such measures will satisfy requirements {\it C1}, {\it C2a}, and {\it C3} \cite{Baumgratz2014}. Additional contractivity requirements are needed for a distance  $D$ in order for the corresponding ${\cal C}_D$ to obey {\it C2b} as well \cite{VedralPlenioPRA}. For instance, while the fidelity-based geometric measure of coherence has been recently proven to be a full coherence monotone \cite{StreltsovSingh}, a related coherence quantifier defined via the squared Bures distance (which is contractive and jointly convex) is known not to  satisfy {\it C2b} \cite{Shao2014}.

All our subsequent findings will apply to bona fide distance-based coherence measures ${\cal C}_D$, which clearly include coherence monotones obeying all the resource-theory requirements recalled earlier. An example of a distance-based coherence monotone is the relative entropy of coherence \cite{Baumgratz2014}, given by
\begin{equation}
{\cal C}_{RE} (\rho) = {\cal S}(\rho_{\rm diag}) - {\cal S}(\rho)
\end{equation}
for any state $\rho$, where $\rho_{\rm diag}$ is the matrix containing only the leading diagonal elements of $\rho$ in the reference basis, and ${\cal S}(\rho) = -\text{Tr}(\rho \log \rho)$ is the von Neumann entropy.

We can also define the trace distance of coherence ${\cal C}_{\text{Tr}}$ as in Eq.~(\ref{eq:CDdef}) using the bona fide trace distance $D_{\text{Tr}}(\rho, \tau) = \frac12 \text{Tr} |\rho-\tau|$. For one-qubit states $\rho$, the trace distance of coherence equals (half) the $l_1$-norm of coherence \cite{Shao2014,Nakano2013}, but this equivalence is not valid for higher dimensional systems, and it is still unknown whether ${\cal C}_{\text{Tr}}$ obeys requirement {\it C2b} in general.

\smallskip
\noindent \emph{\bfseries Frozen coherence: one qubit.} We now analyse conditions such that the  $l_{1}$-norm and relative entropy of coherence are invariant during the evolution of a single qubit (initially in a state $\rho$) under any of the noisy channels $\Lambda_q^\Xi$ described above. This is done by imposing a vanishing differential of the measures  on the evolved state, $\partial_q {\cal C}(\Lambda_q^\Xi(\rho))=0\ \ \forall q\in[0,1]$, with respect to the noise parameter $q$, which can also be interpreted as a dimensionless time \cite{notetime}. We find that only the bit and bit-phase flip channels allow for nonzero frozen coherence (in the computational basis), while all the other considered incoherent channels leave coherence invariant only trivially when the initial state is already incoherent. We can then ask whether nontrivial common freezing conditions for ${\cal C}_{l_1}$ and ${\cal C}_{RE}$ exist.

Writing a single-qubit state in general as  $\rho = \frac{1}{2}(\mathbb{I}+\sum_{j} n_{j} \sigma_{j})$ in terms of its Bloch vector $\vec{n} = \{n_1, n_2, n_3\}$, the bit flip channel $\Lambda_q^{F_1}$ maps an initial Bloch vector $\vec{n}(0)$ to an evolved one $\vec{n}(q) = \{n_{1}(0),(1-q)  n_{2}(0),(1-q)   n_{3}(0)\}$. As the $l_1$-norm of coherence is independent of $n_3$, while $n_1$ is unaffected by the channel, we get that necessary and sufficient freezing conditions for ${\cal C}_{l_1}$ under a single-qubit bit flip channel amount to $n_2(0)=0$ in the initial state. Similar conclusions apply to the bit-phase flip channel $\Lambda_q^{F_2}$ by swapping the roles of $n_1$ and $n_2$.

Conversely, the relative entropy of coherence is also dependent on  $n_{3}$. By analysing the $q$-derivative of ${\cal C}_{RE}$, we see that such a measure is frozen through the bit flip channel only when either $n_{1}(0)=0$ and $n_{2}(0)=0$ (trivial because the initial state is incoherent) or  $n_{2}(0)=0$ and $n_{3}(0)=0$ (trivial because the initial state is invariant under the channel).
Therefore, there is no nontrivial freezing of the relative entropy of coherence under the bit flip or bit-phase flip channel either.

We conclude that, although the $l_1$-norm of coherence can be frozen for specific initial states under flip channels, nontrivial universal freezing of coherence is impossible for the dynamics of a single qubit under paradigmatic incoherent maps.

\smallskip
\noindent \emph{\bfseries Frozen coherence: two qubits.} This is not  true anymore when considering more than one qubit. We will now show that any bona fide distance-based measure of quantum coherence manifests freezing forever in the case of two qubits $A$ and $B$ undergoing local identical bit flip channels \cite{noteflip} and starting from the  initial conditions specified as follows. We consider two-qubit states with maximally mixed marginals (\BD states), also known as Bell-diagonal states \cite{HoroBell}, which are identified by a triple $\vec{c}=\{c_1,c_2,c_3\}$ in their Bloch representation
\begin{equation}
\rho = \frac{1}{4}\left( \mathbb{I}^A\otimes\mathbb{I}^B + {\sum}_{j=1}^3 c_j \sigma_j^A\otimes\sigma_j^B \right).
\end{equation}
Local bit flip channels on each qubit map initial \BD states with $\vec{c}(0) = \{c_1(0), c_2(0), c_3(0)\}$ to \BD states with $\vec{c}(q) = \{c_1(0), (1-q)^2 c_2(0), (1-q)^2 c_3(0)\}$.
Then, the subset of \BD states supporting frozen  coherence for all bona fide distance-based measures is given by the initial condition  \cite{Mazzola2010, Aaronson2013, Cianciaruso2014},
\begin{equation}\label{eq:frozen2}
c_2(0) = - c_1(0) c_3(0)\,.
\end{equation}

To establish this claim, we first enunciate   two auxiliary results, which simplify the evaluation of distance-based coherence monotones  (\ref{eq:CDdef}) for the relevant class of \BD states.

\noindent \textit{Lemma 1.} According to any contractive and convex distance $D$, one of the closest incoherent states $\delta_\rho$ to a \BD state $\rho$ is always a \BD incoherent state, i.e.~one of the form
\begin{equation}\label{eq:closestincoherentstatetoaBDstate}
\delta_\rho = \mbox{$\frac{1}{4}\left(\mathbb{I}^A\otimes\mathbb{I}^B + s
\sigma_3^A\otimes\sigma_3^B \right)$\,, for some $s\in[-1,1]$}\,.
\end{equation}
\noindent \textit{Lemma 2.} According to any contractive and convex distance $D$, one of the closest incoherent states $\delta_\rho$ to a \BD state $\rho$ with triple $\{c_1,-c_1c_3,c_3\}$ is the \BD state $\delta_\rho$ with triple $\{0,0,c_3\}$.

It then follows  that any bona fide distance-based measure of  coherence ${\cal C}_{D}$ for the  \BD states $\rho(q)$, evolving from the initial conditions (\ref{eq:frozen2}) under local bit flip channels, is given by
\begin{eqnarray}\label{eq:frozenquantumcoherence}
{\cal C}_{D}(\rho(q))&=&D\big(\{c_1(0),-(1-q)^2 c_1(0) c_3(0), (1-q)^2 c_3(0)\}, \nonumber \\ & &
\quad\ \{0,0,(1-q)^2c_3(0)\}\big)  \,\,=\,\, {\cal C}_D(\rho(0))\,,\nonumber
\end{eqnarray}
which is frozen for any $q \in [0,1]$, or equivalently frozen forever for any $t$ \cite{notetime}.
The two Lemmas and the main implication on frozen coherence can be rigorously proven by invoking and adapting recent results on the dynamics of quantum correlations for \BD states, reported in~\cite{Cianciaruso2014}. A comprehensive proof is provided in the Supplemental Material \cite{epaps}. This finding shows that, in contrast to the one-qubit case, universal freezing of quantum coherence---measured within a bona fide geometric approach---can in fact occur in two-qubit systems exposed to conventional local decohering dynamics.

Coming back now to the two specific coherence monotones analysed here \cite{Baumgratz2014}, we know that the relative entropy of coherence ${\cal C}_{RE}$ is a bona fide distance-based measure, hence it manifests freezing in the conditions of Eq.~(\ref{eq:frozen2}). Interestingly, we will now show that the  $l_1$-norm of coherence ${\cal C}_{l_1}$ coincides with (twice) the trace distance of coherence ${\cal C}_{\text{Tr}}$ for any \BD state, which implies that ${\cal C}_{l_1}$ also freezes in the same dynamical conditions. To this aim we need to show that, with respect to the trace distance $D_{\text{Tr}}$, one of the closest incoherent states $\delta_\rho$ to a \BD state $\rho$ is always its diagonal part $\rho_{\rm diag}$. The trace distance between a \BD state $\rho$ with $\{c_1,c_2,c_3\}$ and one of its closest incoherent states $\delta_\rho$, which is itself a \BD state of the form (\ref{eq:closestincoherentstatetoaBDstate}) according to Lemma 1, is given by $D_{\text{Tr}}(\rho,\delta_\rho)=\frac{1}{4}(|s + c_1 - c_2 - c_3| + |s - c_1 + c_2 - c_3| + |s + c_1 + c_2 - c_3| + |-s + c_1 + c_2 + c_3|)$. It is immediate to see that the minimum over $\delta_\rho$ is attained by $s=c_3$, i.e., by $\delta_\rho=\rho_{\rm diag}$ as claimed. Notice, however, that
the equivalence between ${\cal C}_{l_1}$ and ${\cal C}_{\text{Tr}}$ does not extend to general two-qubit states, as can be confirmed numerically.

Similarly to the single-qubit case, we can derive a larger set of necessary and sufficient freezing conditions valid specifically for the $l_1$-norm of coherence. Every two-qubit state $\rho$ can be transformed, by local unitaries, into a standard form \cite{Luo2008} with Bloch representation
$\rho = \frac{1}{4}\big( \mathbb{I}^A\otimes\mathbb{I}^B + \sum_{j=1}^3 x_{j} \sigma_{j}^{A} \otimes \mathbb{I}^{B} + \sum_{j=1}^3 y_{j} \mathbb{I}^{A} \otimes \sigma_{j}^{B} + \sum_{j=1}^3 T_{jj} \sigma_j^A\otimes\sigma_j^B \big)$.
We have then that initial states of this form, with $x_1$, $y_1$, $x_3$, $y_3$, $T_{33}$ arbitrary, $x_2=y_2=0$, and $T_{22}=u T_{11}$ with $u \in [-1,1]$, manifest frozen coherence as measured by ${\cal C}_{l_1}$ under local bit flip channels; however, the same does not hold for ${\cal C}_{RE}$ in general.

\smallskip
\noindent \emph{\bfseries Frozen coherence: N qubits.} Our main finding can be readily generalised to a system of $N$ qubits with any even $N$. We define $N$-qubit states with maximally mixed marginals (\NBD states) \cite{Xu2013,Chanda2014} as those with density matrix of the form
$\rho = \frac{1}{2^{N}} \left( \mathbb{I}^{\otimes N} + {\sum}_{j=1}^{3} c_{j} \sigma_{j}^{\otimes N}\right)$,
still specified by the triple $\{c_1, c_2, c_3\}$ as in the $N=2$ case.
We have then that, when the system is evolving according to identical and independent local bit flip channels acting on each qubit as in  Eq.~(\ref{eq:independentchannels}) with $\Xi = F_1$, the quantum coherence of the system is universally frozen according to any bona fide distance-based measure if the $N$ qubits are initialised in a \NBD state respecting the freezing condition
\begin{equation}\label{eq:frozenN}
c_2(0) = (-1)^{N/2} c_1(0) c_3(0)\,,
\end{equation}
which generalises (\ref{eq:frozen2}). This is the most general result of the present Letter \cite{noteflip}, and its full proof is provided in the Supplemental Material \cite{epaps}.
We observe that, by virtue of the formal equivalence between a system of $N$ qubits  and a single qudit with dimension $d=2^N$, our results can also be interpreted as providing universal freezing conditions for all bona fide distance-based measures of coherence in a single $2^N$-dimensional system with any even $N$. Naturally, one may expect larger sets of freezing conditions to exist for specific coherence monotones such as the $l_1$-norm, like in the $N=2$ case; their characterisation is outside the scope of this Letter.

We further note that no universal freezing of coherence is instead possible for  \NBD states with odd $N$, whose dynamical properties are totally analogous to those of one-qubit states.

\smallskip
\noindent \emph{\bfseries Coherence versus quantum correlations.} The freezing conditions established here for coherence have been in fact identified in previous literature \cite{Maziero2009, Mazzola2010, Aaronson2013, Cianciaruso2014, Chanda2014}, as various measures of so-called discord-type quantum correlations  were shown to freeze under the same dynamical conditions up to a threshold time $t^\star$, defined in our notation \cite{notetime} by the largest value of $q$ such that $|c_3(q)| \geq |c_1(q)|$, for \NBD states evolving under local bit flip channels. Focusing on the two-qubit case for clarity, we note that for \BD states with $|c_3| \geq |c_1|$, and for any bona fide distance $D$, the distance-based measure of coherence ${\cal C}_D$, defined by Eq.~(\ref{eq:CDdef}) and evaluated in Eq.~(\ref{eq:frozenquantumcoherence}), {\it coincides} with the corresponding distance-based measure of discord-type quantum correlations ${\cal Q}_D$, formalised e.g.~in Ref.~\cite{Cianciaruso2014}. Hence, the freezing of coherence might provide a deeper insight into the peculiar phenomenon of frozen quantum correlations under local flip channels (see also \cite{Banik2014}), as the latter just reduce to coherence for $t \leq t^\star$  under the conditions we identified.

More generally, measures of discord-type correlations \cite{Modi2012,Streltsov2015,interpower} may be  recast as suitable measures of coherence in bipartite systems, minimised over the reference basis, with minimisation restricted to local product bases. For instance, the minimum $l_1$-norm of coherence \cite{Baumgratz2014} yields the negativity of quantumness \cite{Piani2011,Nakano2013,Sciarrino2014}, the minimum relative entropy of coherence \cite{Baumgratz2014} yields the relative entropy of discord \cite{Horodecki2005,Modi2010,Piani2011,Streltsov2011}, and the minimum skew information \cite{Girolami2014} yields the local quantum uncertainty \cite{lqu}. Our result suggest therefore that the computational basis is the product basis which minimises coherence (according to suitable bona fide measures) for particular \BD states undergoing local bit flip noise $\Lambda^{F_1}$ up to $t \leq t^\star$, while coherence is afterwards minimised in the eigenbasis of $\sigma_1$, which is the pointer basis towards which the system eventually converges due to the local decoherence \cite{Cornelio2012}; similar conclusions can be drawn for the other $k$-flip channels \cite{noteflip}.

We finally remark that, unlike more general discord-type correlations,  entanglement \cite{Horodecki2009} plays no special role in the freezing phenomenon analysed in this Letter, as the latter can also happen for states that remain separable during the whole evolution, e.g.~the \BD states with initial triple $\{\frac{1}{4},-\frac{1}{16},\frac{1}{4}\}$.

\smallskip
\noindent \emph{\bfseries Conclusions.} We have determined exact conditions such that any bona fide distance-based measure of quantum coherence \cite{Baumgratz2014}
is dynamically frozen:  this occurs for an even number of qubits, initialised in a particular class of states with maximally mixed marginals, and undergoing local independent and identical nondissipative flip channels (Fig.~\ref{frolla}). We have also shown that there is no general agreement on freezing conditions between specific coherence monotones when considering either the one-qubit case or more general $N$-qubit initial states.  This highlights the prominent role played by the aforementioned universal freezing conditions in ensuring a durable physical exploitation of  coherence, regardless of how it is quantified, for applications such as quantum metrology \cite{Giovannetti2004} and nanoscale thermodynamics \cite{Aaberg2014,Lostaglio2014}.
It will be interesting to explore practical realisations of such dynamical conditions  \cite{Xu2010,Auccaise2011,Cornelio2012,xulofranco2013NatComms,Silva2013,PaulaPRL}.

Complex systems are inevitably subject to noise, hence it is natural and technologically crucial to question under what conditions the quantum resources that we can extract from them are not deteriorated during open evolutions \cite{andreascoh}. In addressing this problem by focusing on coherence, we have also revealed an intrinsic physical explanation for the freezing of discord-type correlations \cite{Cianciaruso2014}, by exposing and exploiting the intimate link between these two nonclassical signatures. Providing unified quantitative resource-theory frameworks for coherence, entanglement, and other quantum correlations is certainly a task worthy of further investigation \cite{StreltsovSingh}.

\noindent \emph{\bfseries Acknowledgements.} We acknowledge fruitful discussions with  E.~Arendelle, A.~Bera, M.~N.~Bera, T.~Chanda, H.~S.~Dhar, A.~Khan, P.~Liuzzo Scorpo, M.~Lock, S.~Luo, A.~K.~Pal, M.~Piani, M.~B.~Plenio, W.~Roga, A.~Sen(De), U.~Sen, I.~A.~Silva, U.~Singh, A.~Streltsov. We thank the ERC StG GQCOP (Grant Agreement No.~637352), the Foundational Questions Institute (Grant No.~FQXi-RFP3-1317), and the Brazilian CAPES (Grant No.~108/2012) for financial support.




\renewcommand\appendixname{Frozen Quantum Coherence: Supplemental Material}

\newtheorem{theorem}{Theorem}[section]
\newtheorem{lemmaA}[theorem]{Lemma A}

\appendix*
\section{Appendix: Supplemental Material}
Consider the $N$-qubit states with the following matrix representation in the computational basis:
\begin{equation}\label{eq:NqubitBellState}
\rho = \frac{1}{2^{N}} \left( \mathbb{I}^{\otimes N} + \sum_{i=1}^{3} c_{i} \sigma_{i}^{\otimes N}\right),
\end{equation}
where $\mathbb{I}$ is the $2 \times 2$ identity matrix, $\sigma_{i}$ is the $i$-th Pauli matrix and $c_{i} = \text{Tr}\left[\rho \sigma_{i}^{\otimes N}\right]\in[-1,1]$. These states will be referred to as \NBD states, as they have maximally mixed marginals (by tracing out any $K < N$ qubits), and will be uniquely identified by the triple $\{c_{1},c_{2},c_{3}\}$.

In this appendix we will show that, for an even number $N$ of qubits, all bona fide distance-based measures of quantum coherence will exhibit the freezing phenomenon when each qubit is subject to local independent bit flip noise, for an initial  \NBD state specified by $\{c_{1},(-1)^{N/2}c_{1}c_{3},c_{3}\}$.

The evolution of an $N$-qubit state $\rho$ under local independent identical $k$-flip channels, where the index  $k\in\{1,2,3\}$ respectively identifies the bit flip ($k=1$), bit-phase flip ($k=2$), and phase flip ($k=3$) channel, can be characterised in the operator-sum representation by the map
\begin{equation}\label{eq:independentbitflipchannels}
\Lambda_{q}^{F_k \otimes N}(\rho) = \sum_{j_1,j_2,\cdots,j_N} K^{F_k}_{j_1}\otimes K^{F_k}_{j_2}\otimes\cdots\otimes K^{F_k}_{j_N} \rho {K^{F_k}_{j_1}}^\dagger\otimes {K^{F_k}_{j_2}}^\dagger\otimes\cdots\otimes {K^{F_k}_{j_N}}^\dagger\,
\end{equation}
where the single-qubit Kraus operators $K_j^{F_k}$ are reported in the main text in terms of the strength of the noise $q \in [0,1]$, which in dynamical terms can be expressed as $q(t)=1-\exp(-\gamma t)$ with $t$ representing time and $\gamma$ being the decoherence rate.
From Eqs.~(\ref{eq:NqubitBellState}) and (\ref{eq:independentbitflipchannels}), one can easily see that $N$ non-interacting qubits initially in a \NBD state, undergoing local identical flip channels, evolve preserving the \NBD structure during the entire dynamics (i.e, for all $q \in [0,1]$, or equivalently for all $t \geq 0$). More precisely, the triple $\{c_1(q),c_2(q),c_3(q)\}$ characterising the \NBD evolved state $\rho(q)$ can be written as follows
\begin{equation}\label{eq:timeevolvedNBDstateunderlocalpuredephasing}
c_{i,j\neq k}(q)=(1-q)^N c_{i,j\neq k}(0),\ \ \ c_k(q)=c_k(0)\,,
\end{equation}
where $\{c_1(0),c_2(0),c_3(0)\}$ is the triple characterising the initial \NBD state $\rho$.

We start by showing that, for even $N$, the eigenvectors and eigenvalues of an arbitrary \NBD state $\rho$ are given by, respectively
\begin{eqnarray}\label{eq:NqubitBellBasis}
|\beta_1^{\pm}\rangle&=&\frac{1}{\sqrt{2}}\left(|000\ldots 000\rangle \pm |111\ldots 111\rangle  \right), \\
|\beta_2^{\pm}\rangle&=&\frac{1}{\sqrt{2}}\left(|000\ldots 001\rangle \pm |111\ldots 110\rangle  \right),\nonumber \\
|\beta_3^{\pm}\rangle&=&\frac{1}{\sqrt{2}}\left(|000\ldots 010\rangle \pm |111\ldots 101\rangle  \right),\nonumber \\
|\beta_4^{\pm}\rangle&=&\frac{1}{\sqrt{2}}\left(|000\ldots 011\rangle \pm |111\ldots 100\rangle  \right), \nonumber \\
\cdots&  & \nonumber \\
|\beta_{2^{N-1}-1}^{\pm}\rangle&=&\frac{1}{\sqrt{2}}\left(|011\ldots 110\rangle \pm |100\ldots 001\rangle  \right),\nonumber \\
|\beta_{2^{N-1}}^{\pm}\rangle&=&\frac{1}{\sqrt{2}}\left(|011\ldots 111\rangle \pm |100\ldots 000\rangle  \right),\nonumber \\ \nonumber
\end{eqnarray}
and
\begin{equation}\label{eq:EigenvaluesNBellStates}
\lambda_p^{\pm}=\frac{1}{2^N}\left[1\pm c_1 \pm (-1)^{N/2}(-1)^p c_2 +(-1)^p c_3 \right],
\end{equation}
where $p$ is the parity of $|\beta_i^{\pm}\rangle$ with respect to the parity operator along the $z$-axis $\Pi_3 \equiv \sigma_3^{\otimes N}$, i.e.
\begin{equation}\label{eq:definitionofp}
\Pi_3 |\beta_i^{\pm}\rangle = (-1)^p |\beta_i^{\pm}\rangle.
\end{equation}

It will suffice to prove the following equation:
\begin{equation}\label{eq:NBDstateEigenSystem}
\rho |\beta_i^{\pm}\rangle = \lambda_p^{\pm} |\beta_i^{\pm}\rangle,
\end{equation}
for any $i\in\{1,\cdots,2^{N-1}\}$. In fact, by writing a generic state $|\beta_i^{\pm}\rangle$ as follows
\begin{eqnarray}
|\beta_i^{\pm}\rangle = \frac{1}{\sqrt{2}} \left(|p,N-p\rangle \pm |N-p,p\rangle \right),
\end{eqnarray}
where $|n_0,n_1\rangle$ denotes any element of the $N$-qubit computational basis whose number of $0$'s ($1$'s) is equal to $n_0$ ($n_1$), one can easily see that
\begin{eqnarray}
\sigma_1^{\otimes N} |\beta_i^{\pm}\rangle &=& \frac{1}{\sqrt{2}} \left(|N-p,p\rangle \pm |p,N-p\rangle \right),\nonumber\\
\sigma_2^{\otimes N} |\beta_i^{\pm}\rangle &=& \frac{1}{\sqrt{2}} (-1)^{N/2} \left[(-1)^{p}|N-p,p\rangle \pm (-1)^{N-p}|p,N-p\rangle \right], \nonumber \\
\sigma_3^{\otimes N} |\beta_i^{\pm}\rangle &=& \frac{1}{\sqrt{2}} \left[(-1)^{N-p}|p,N-p\rangle \pm (-1)^{p}|N-p,p\rangle \right]. \nonumber
\end{eqnarray}
Eventually, by using the above three equations, Eq. (\ref{eq:NqubitBellState}) and the fact that $N$ is even, so that $(-1)^{N-p}=(-1)^p$, one can easily verify that $\rho|\beta_i^{\pm}\rangle$ is equal to $ \lambda_p^{\pm} |\beta_i^{\pm}\rangle$, i.e.~that Eq. (\ref{eq:NBDstateEigenSystem}) holds.

Now we are ready to show the three essential pieces which will lead us to prove the main result on the universal freezing phenomenon of bona fide distance-based measures of quantum coherence in the $N$-qubit setting (with even $N$).

\begin{lemmaA}\label{th:anycontractivedistanceismagictranslationalinvariant}
For all even $N$, any contractive distance satisfies the following translational invariance properties within the space of $N$-qubit \NBD states:
\begin{equation}\label{eq:magictranslationalinvariance1}
D(\{c_1,(-1)^{N/2}c_1c_3,c_3\},\{c_1,0,0\}) = D(\{0,0,c_3\},\{0,0,0\})
\end{equation}
and
\begin{equation}\label{eq:magictranslationalinvariance2}
D(\{c_1,(-1)^{N/2}c_1c_3,c_3\},\{0,0,c_3\}) = D(\{c_1,0,0\},\{0,0,0\})
\end{equation}for all $c_1$ and $c_3$, where $\{c_1,(-1)^{N/2}c_1c_3,c_3\}$ denotes a \NBD state in Eq. (\ref{eq:NqubitBellState}) with $c_2=(-1)^{N/2}c_1c_3$.
\end{lemmaA}

\noindent \textit{Proof}.
Let us start by proving Eq.~(\ref{eq:magictranslationalinvariance1}). First of all, by considering the channel $\Lambda_1^{F_3 \otimes N}$ representing the local independent phase flip noise expressed by Eq.~(\ref{eq:independentbitflipchannels}), when $k=3$ and $q=1$  (i.e.~$t \rightarrow \infty $), we have the following inequality
\begin{eqnarray} \label{eq:inequalityduetodephasing1}
\!\!\!\!\!\!\!\!\!\!\!\!\!\!\!&&D(\{0,0,c_3\},\{0,0,0\}) \nonumber \\
\!\!\!\!\!\!\!\!\!\!\!\!\!\!\!&=& D(\Lambda_1^{F_3 \otimes N}\{c_1,(-1)^{N/2}c_1 c_3,c_3\},\Lambda_1^{F_3 \otimes N}\{c_1,0,0\})   \\
\!\!\!\!\!\!\!\!\!\!\!\!\!\!\!&\leq& D(\{c_1,(-1)^{N/2}c_1 c_3,c_3\},\{c_1,0,0\}), \nonumber
\end{eqnarray}
where the first equality is due to the fact that
\begin{eqnarray}
\{0,0,c_3\} &=& \Lambda_1^{F_3 \otimes N} \{c_1,(-1)^{N/2}c_1 c_3,c_3\}, \mbox{ and}\\
\{0,0,0\} &=& \Lambda_1^{F_3 \otimes N}\{c_1,0,0\},
\end{eqnarray}
while the final inequality in (\ref{eq:inequalityduetodephasing1}) is due to the contractivity of the distance $D$.

In order to prove the opposite inequality and thus Eq.~(\ref{eq:magictranslationalinvariance1}), we now introduce a $N$-qubit global {\it rephasing} channel $\Lambda^{R_3}_{r}$ which is defined in the operator-sum representation as
\begin{equation}\label{Eq:KrausRepresentation}
\Lambda^{R_3}_{r}(\rho) = \sum_{i,\pm} K^{R_3}_{i,\pm}\rho {K^{R_3}_{i,\pm}}^{\dagger}\,,
\end{equation}
with
\begin{eqnarray}\label{Eq:KrausOperators}
K^{R_3}_{1,\pm} &=& \sqrt{\frac{1\pm r}{2}}|\beta_1^{\pm}\rangle\langle 000\ldots 000|,\\
K^{R_3}_{2,\pm}  &=& \sqrt{\frac{1\pm r}{2}} |\beta_2^{\pm}\rangle\langle 000\ldots 001|,\nonumber\\
K^{R_3}_{3,\pm}  &=& \sqrt{\frac{1\pm r}{2}} |\beta_3^{\pm}\rangle\langle 000\ldots 010|,\nonumber\\
K^{R_3}_{4,\pm}  &=& \sqrt{\frac{1\pm r}{2}} |\beta_4^{\pm}\rangle\langle 000\ldots 011|,\nonumber\\
\cdots\nonumber&&\\
K^{R_3}_{2^{N-1}-1,\pm}  &=&  \sqrt{\frac{1\pm r}{2}}|\beta_{2^{N-1}-1}^{\pm}\rangle\langle 011\ldots 110|,\nonumber\\
K^{R_3}_{2^{N-1},\pm}  &=&  \sqrt{\frac{1\pm r}{2}}|\beta_{2^{N-1}}^{\pm}\rangle\langle 011\ldots 111|,\nonumber\\
K^{R_3}_{2^{N-1}+1,\pm}  &=& \sqrt{\frac{1\pm r}{2}} |\beta_{2^{N-1}}^{\pm}\rangle\langle 100\ldots 000|,\nonumber\\
K^{R_3}_{2^{N-1}+2,\pm}  &=& \sqrt{\frac{1\pm r}{2}} |\beta_{2^{N-1}-1}^{\pm}\rangle\langle 100\ldots 001|,\nonumber\\
\cdots\nonumber&&\\
K^{R_3}_{2^{N}-3,\pm}  &=& \sqrt{\frac{1\pm r}{2}} |\beta_{4}^{\pm}\rangle\langle 111\ldots 100|,\nonumber\\
K^{R_3}_{2^{N}-2,\pm}  &=& \sqrt{\frac{1\pm r}{2}} |\beta_{3}^{\pm}\rangle\langle 111\ldots 101|,\nonumber\\
K^{R_3}_{2^{N}-1,\pm}  &=& \sqrt{\frac{1\pm r}{2}} |\beta_{2}^{\pm}\rangle\langle 111\ldots 110|,\nonumber\\
K^{R_3}_{2^{N},\pm}  &=& \sqrt{\frac{1\pm r}{2}} |\beta_{1}^{\pm}\rangle\langle 111\ldots 111|,\nonumber
\end{eqnarray}
where $r \in [0,1]$ is a parameter denoting the rephasing strength, $\{ |\beta_i^{\pm}\rangle\}$ is the $N$-qubit basis defined in Eq.~(\ref{eq:NqubitBellBasis}), and the $2^{N+1}$ Kraus operators satisfy $\sum_{i,\pm} {K^{R_3}_{i,\pm}}^{\dagger} K^{R_3}_{i,\pm} = \mathbb{I}^{\otimes N}$, thus ensuring that $\Lambda^{R_3}_{r}$ is a CPTP map.

It is now essential to see that the effect of $\Lambda^{R_3}_{r}$ on a \NBD state of the form $\{0,0,c_3\}$ is given by
\begin{equation}\label{eq:ActionoftheglobalNqubitrephasingchannel}
\Lambda^{R_3}_{r}(\{0,0,c_3\})=\{r,(-1)^{N/2} r \  c_3,c_3\},
\end{equation}
for any even $N$. To prove Eq.~(\ref{eq:ActionoftheglobalNqubitrephasingchannel}), it will be useful to split the $N$-qubit states $|\beta_i^\pm\rangle$ into the states $|\Phi_i^{\pm}\rangle$ and $|\Psi_i^{\pm}\rangle$ with even and odd parity, respectively, i.e. such that
\begin{eqnarray}
\Pi_3 |\Phi_i^{\pm}\rangle&=&|\Phi_i^{\pm}\rangle,\nonumber\\
\Pi_3 |\Psi_i^{\pm}\rangle&=&-|\Psi_i^{\pm}\rangle, \label{eq:PhiandPsi}
\end{eqnarray}
where $i\in\{1,\cdots,2^{N-2}\}$. Thanks to Eqs. (\ref{eq:EigenvaluesNBellStates}), (\ref{eq:definitionofp}) (\ref{eq:NBDstateEigenSystem}) and (\ref{eq:PhiandPsi}), one gets that the spectral decomposition of a \NBD state $\rho_{\{c_1,c_2,c_3\}}$ with generic triple $\{c_1,c_2,c_3\}$ can be written as follows,
\begin{eqnarray}
\rho_{\{c_1,c_2,c_3\}} \\ \nonumber
&=& \frac{1}{2^N}\left[1 + c_1 + (-1)^{N/2} c_2 + c_3 \right]\sum_i|\Phi_i^{+}\rangle\langle\Phi_i^{+}|\\ \nonumber
&+& \frac{1}{2^N}\left[1 - c_1 - (-1)^{N/2} c_2 + c_3 \right]\sum_i|\Phi_i^{-}\rangle\langle\Phi_i^{-}| \\ \nonumber
&+&\frac{1}{2^N}\left[1 + c_1 - (-1)^{N/2} c_2 - c_3 \right]\sum_i|\Psi_i^{+}\rangle\langle\Psi_i^{+}|\\ \nonumber
&+& \frac{1}{2^N}\left[1 - c_1 + (-1)^{N/2}c_2 - c_3 \right] \sum_i|\Psi_i^{-}\rangle\langle\Psi_i^{-}|
\end{eqnarray}
As a consequence
\begin{eqnarray}
\rho_{\{0,0,c_3\}} \\  \nonumber
&=& \frac{1}{2^N}\left(1 + c_3 \right)\sum_i|\Phi_i^{+}\rangle\langle\Phi_i^{+}|\\ \nonumber
&+& \frac{1}{2^N}\left(1 + c_3 \right)\sum_i|\Phi_i^{-}\rangle\langle\Phi_i^{-}|\\ \nonumber
&+&\frac{1}{2^N}\left(1 - c_3 \right)\sum_i|\Psi_i^{+}\rangle\langle\Psi_i^{+}|\\ \nonumber
&+&\frac{1}{2^N}\left(1 - c_3 \right)\sum_i|\Psi_i^{-}\rangle\langle\Psi_i^{-}|,
\end{eqnarray}
while
\begin{eqnarray}
\rho_{\{r,(-1)^{N/2} r \  c_3,c_3\}}  \\ \nonumber
&=& \frac{1}{2^N}(1+r)(1+c_3)\sum_i|\Phi_i^{+}\rangle\langle\Phi_i^{+}|\\ \nonumber
&+& \frac{1}{2^N}(1-r)(1+c_3)\sum_i|\Phi_i^{-}\rangle\langle\Phi_i^{-}| \\ \nonumber
&+& \frac{1}{2^N}(1+r)(1-c_3)\sum_i|\Psi_i^{+}\rangle\langle\Psi_i^{+}|\\ \nonumber
&+& \frac{1}{2^N}(1-r)(1-c_3) \sum_i|\Psi_i^{-}\rangle\langle\Psi_i^{-}|. \\ \nonumber
\end{eqnarray}
By exploiting the following equalities
\begin{eqnarray}
\Lambda^{R_3}_{r}(|\Phi_i^{+}\rangle\langle\Phi_i^{+}|)=\frac{1+r}{2}|\Phi_i^{+}\rangle\langle\Phi_i^{+}|+\frac{1-r}{2}|\Phi_i^{-}\rangle\langle\Phi_i^{-}|,\\ \nonumber
\Lambda^{R_3}_{r}(|\Phi_i^{-}\rangle\langle\Phi_i^{-}|)=\frac{1+r}{2}|\Phi_i^{+}\rangle\langle\Phi_i^{+}|+\frac{1-r}{2}|\Phi_i^{-}\rangle\langle\Phi_i^{-}|,\\ \nonumber\\ \nonumber
\Lambda^{R_3}_{r}(|\Psi_i^{+}\rangle\langle\Psi_i^{+}|)=\frac{1+r}{2}|\Psi_i^{+}\rangle\langle\Psi_i^{+}|+\frac{1-r}{2}|\Psi_i^{-}\rangle\langle\Psi_i^{-}|,\\ \nonumber\\ \nonumber
\Lambda^{R_3}_{r}(|\Psi_i^{-}\rangle\langle\Psi_i^{-}|)=\frac{1+r}{2}|\Psi_i^{+}\rangle\langle\Psi_i^{+}|+\frac{1-r}{2}|\Psi_i^{-}\rangle\langle\Psi_i^{-}|,
\end{eqnarray}
and the linearity of the global rephasing channel, we get
\begin{eqnarray}
\Lambda^{R_3}_{r}(\{0,0,c_3\})
&=& \frac{1}{2^N}\left(1 + c_3 \right)\sum_i\Lambda^{R_3}_{r}(|\Phi_i^{+}\rangle\langle\Phi_i^{+}|)\\ \nonumber
&+& \frac{1}{2^N}\left(1 + c_3 \right)\sum_i\Lambda^{R_3}_{r}(|\Phi_i^{-}\rangle\langle\Phi_i^{-}|)\\ \nonumber
&+&\frac{1}{2^N}\left(1 - c_3 \right)\sum_i\Lambda^{R_3}_{r}(|\Psi_i^{+}\rangle\langle\Psi_i^{+}|)\\ \nonumber
&+&\frac{1}{2^N}\left(1 - c_3 \right)\sum_i\Lambda^{R_3}_{r}(|\Psi_i^{-}\rangle\langle\Psi_i^{-}|)\\ \nonumber
&=& \{r,(-1)^{N/2} r \  c_3,c_3\}. \nonumber
\end{eqnarray}

We then have the inequality
\begin{eqnarray}\label{eq:inequalityduetorephasing1}
D(\{c_1,(-1)^{N/2}c_1 c_3,c_3\},\{c_1,0,0\})\nonumber \\
= D(\Lambda^{R_3}_{c_1}\{0,0,c_3\},\Lambda^{R_3}_{c_1}\{0,0,0\})   \\
\leq D(\{0,0,c_3\},\{0,0,0\}),\nonumber
\end{eqnarray}
where the first equality is due to the fact that
\begin{eqnarray}
\{c_1,(-1)^{N/2}c_1 c_3,c_3\} &=&\Lambda^{R_3}_{c_1}\{0,0,c_3\},\mbox{ and}\nonumber\\
\{c_1,0,0\} &=& \Lambda^{R_3}_{c_1}\{0,0,0\},\nonumber
\end{eqnarray}
while the final inequality in (\ref{eq:inequalityduetorephasing1}) is again due to the contractivity of the distance $D$. By putting together the two inequalities (\ref{eq:inequalityduetodephasing1}) and (\ref{eq:inequalityduetorephasing1}), we immediately get the invariance of  Eq.~(\ref{eq:magictranslationalinvariance1}) for any contractive distance.

In order now to prove Eq.~(\ref{eq:magictranslationalinvariance2}), we introduce the local unitary $V^{\otimes N}$ with $V=\frac{1}{\sqrt{2}}(\mathbb{I}+i\sigma_2)$. The effect of $V^{\otimes N}$ on a general \NBD state is given by
\begin{equation}\label{eq:UnitarySwappingc1andc3}
V^{\otimes N}\{c_1,c_2,c_3\}{V^{\otimes N}}^\dagger=\{c_3,c_2,c_1\},
\end{equation}
where this can be easily seen by utilising the fact that $N$ is even and the following single-qubit identities:
\begin{eqnarray}
V \sigma_1 V^\dagger &=& \sigma_3,\nonumber \\
V \sigma_2 V^\dagger &=& \sigma_2,\nonumber \\
V \sigma_3 V^\dagger &=& - \sigma_1.\nonumber
\end{eqnarray}
Thanks to the invariance under unitaries of any contractive distance $D$, the effect of the unitary ${V^{\otimes N}}$ expressed by Eq.~(\ref{eq:UnitarySwappingc1andc3}), and the just proven invariance expressed by Eq.~(\ref{eq:magictranslationalinvariance1}), we eventually have
\begin{eqnarray}
& &D(\{c_1,(-1)^{N/2}c_1c_3,c_3\},\{0,0,c_3\})  \\
&=& D({V^{\otimes N}}\{c_1,(-1)^{N/2}c_1c_3,c_3\}{V^{\otimes N}}^\dagger,{V^{\otimes N}}\{0,0,c_3\}{V^{\otimes N}}^\dagger) \nonumber \\
&=& D(\{c_3,(-1)^{N/2}c_1c_3,c_1\},\{c_3,0,0\}) \nonumber \\
&=& D(\{0,0,c_1\},\{0,0,0\}) \nonumber \\
&=& D({V^{\otimes N}}\{0,0,c_1\}{V^{\otimes N}}^\dagger,{V^{\otimes N}}\{0,0,0\}{V^{\otimes N}}^\dagger) \nonumber \\
&=&D(\{c_1,0,0\},\{0,0,0\}),\nonumber
\end{eqnarray}
that is Eq.~(\ref{eq:magictranslationalinvariance2}).
\begin{flushright}
$\blacksquare$
\end{flushright}

\begin{lemmaA}\label{th:nearestincoherentstatetoaNBDstate}
For all even $N$, according to any contractive and convex distance $D$, one of the closest incoherent states $\delta_{\rho}$ to a  \NBD state $\rho$ is always a \NBD  incoherent state, i.e.~one of the form
\begin{equation}\label{eq:closestincoherentstatetoaBDstate}
\delta_\rho=\frac{1}{2^{N}} \left( \mathbb{I}^{\otimes N} + s \  \sigma_{3}^{\otimes N}\right)
\end{equation}
for some coefficient $s\in[-1,1]$.
\end{lemmaA}

\noindent \textit{Proof}.
Consider an arbitrary $N$-qubit state $\rho$, which can be represented as
\begin{equation}\label{eq:correlationmatrixNqubit}
\rho=\frac{1}{2^N}\sum_{i_{1}, i_{2}, \ldots, i_{N}=0}^{3} \tau_{i_{1}i_{2} \ldots i_{N}} \sigma_{i_{1}} \otimes \sigma_{i_{2}} \ldots \otimes \sigma_{i_{N}},
\end{equation}
where the coefficients $\tau_{i_{1},i_{2}, \ldots i_{N}}=\text{Tr}\left[\rho\  \sigma_{i_{1}} \otimes \sigma_{i_{2}} \ldots \otimes \sigma_{i_{N}} \right] \in [-1,1]$ are the correlation tensor elements of $\rho$, and $\sigma_{0} \equiv \mathbb{I}$. Any term involving $\sigma_{1}$ or $\sigma_{2}$ in the tensorial sum (\ref{eq:correlationmatrixNqubit})  introduces off-diagonal elements, therefore we can write a general $N$-qubit incoherent state, with respect to the computational basis, as
\begin{equation}
\delta=\frac{1}{2^N}\sum_{i_{1},i_{2}, \ldots, i_{N}=\{0,3\}}  \tau_{i_{1}i_{2} \ldots i_{N}} \sigma_{i_{1}} \otimes \sigma_{i_{2}} \ldots \otimes \sigma_{i_{N}},
\end{equation}
where each index $i_j$ can now take either $0$ or $3$ as the only values.
For any $N$-qubit incoherent state $\delta$, we can define a corresponding incoherent \NBD state $\delta_{{\rm{M}}^3_N}$, whose $\tau$ tensor is obtained from the one of $\delta$ by setting all the $\tau_{i_{1},i_{2}, \ldots i_{N}}$ equal to zero, but for the two entries $\tau_{00 \ldots 0}$ and $\tau_{33 \ldots 3}$. We want to show that $D(\rho,\delta_{{\rm{M}}^3_N}) \leq D(\rho,\delta)$ for any \NBD state $\rho$ and $N$-qubit incoherent state $\delta$, which readily implies that one of the closest incoherent states $\delta_{\rho}$ to a \NBD state $\rho$ is indeed a \NBD state.

To begin, first consider the family of $N-1$ unitaries
\begin{eqnarray}
\{U_{j}\}_{j=1}^{N-1}=\{(\sigma_{1} \otimes \sigma_{1} \otimes \mathbb{I}^{\otimes N-2}),
(\mathbb{I} \otimes \sigma_{1} \otimes \sigma_{1} \otimes \mathbb{I}^{\otimes N-3}), \\
(\mathbb{I}^{\otimes 2} \otimes \sigma_{1} \otimes \sigma_{1} \otimes \mathbb{I}^{\otimes N-4}),
(\mathbb{I}^{\otimes 3} \otimes \sigma_{1} \otimes \sigma_{1} \otimes \mathbb{I}^{\otimes N-5}),\nonumber \\
\ldots,
(\mathbb{I}^{\otimes N-3} \otimes \sigma_{1} \otimes \sigma_{1} \otimes \mathbb{I}),
(\mathbb{I}^{\otimes N-2} \otimes \sigma_{1} \otimes \sigma_{1})
\}. \nonumber
\end{eqnarray}
We note that every \NBD state $\rho$ is invariant under the action of any $U_{j}$. This can be seen as follows
\begin{eqnarray}
U_{j} \rho U_{j}^{\dagger} = \frac{1}{2^{N}} \left( U_{j} \mathbb{I}^{\otimes N} U_{j}^{\dagger}+ \sum_{i=1}^{3} c_{i} U_{j} \sigma_{i}^{\otimes N} U_{j}^{\dagger} \right) \nonumber \\
=\frac{1}{2^{N}} \left( \mathbb{I}^{\otimes N}+ \sum_{i=1}^{3} c_{i} \sigma_{i}^{\otimes N} \right) = \rho ,
\end{eqnarray}
where in the second equality we use $U_{j} \mathbb{I}^{\otimes N} U_{j}^{\dagger}=\mathbb{I}^{\otimes N}$ and $U_{j} \sigma_{i}^{\otimes N}U_{j}^{\dagger}=\sigma_{i}^{\otimes N}$ which arises simply by recalling $\sigma_{1} \sigma_{1}\sigma_{1}=\sigma_{1}$, $\sigma_{1} \sigma_{2}\sigma_{1}=-\sigma_{2}$ and $\sigma_{1} \sigma_{3}\sigma_{1}=-\sigma_{3}$ and noting that there are always two $\sigma_{1}$'s in each unitary.

Now consider the action of $U_{1}$ on a generic incoherent state $\delta$. The state transforms as
\begin{equation}
U_{1} \delta U_{1}^{\dagger} = \frac{1}{2^N} \!\!\! \sum_{i_{1},i_{2}, \ldots, i_{N}=\{0,3\}}\!\!\!  \tau_{i_{1}i_{2} \ldots i_{N}} \sigma_{1}\sigma_{i_{1}}\sigma_{1} \otimes \sigma_{1}\sigma_{i_{2}}\sigma_{1} \otimes \sigma_{i_{3}}\otimes \ldots \otimes \sigma_{i_{N}}.
\end{equation}
We have $\sigma_{1} \sigma_0 \sigma_{1} = \sigma_0$ and $\sigma_{1} \sigma_{3} \sigma_{1} = - \sigma_{3}$, hence the coefficients $\tau_{i_{1}i_{2} \ldots i_{N}}^{U_{1}}$ of $U_{1} \delta U_{1}^{\dagger}$ are $\tau_{00 \ldots}^{U_{1}}= \tau_{00 \ldots}$, $\tau_{33 \ldots}^{U_{1}}= \tau_{33 \ldots}$, $\tau_{03 \ldots}^{U_{1}}= -\tau_{03 \ldots}$ and $\tau_{30 \ldots}^{U_{1}}= -\tau_{30 \ldots}$;   in other words, $U_{1}$ flips the sign of any element $\tau_{i_{1}i_{2} \ldots i_{N}}$ for which $i_{1} \neq i_{2}$. We can further define a state that is a linear combination of $\delta$ and $U_{1} \delta U_{1}^{\dagger}$,
\begin{equation}
\delta^{1}=\frac{1}{2}(\delta+U_{1} \delta U_{1}^{\dagger}).
\end{equation}
The coefficients $\tau_{i_{1}i_{2} \ldots i_{N}}^{1}$ of $\delta^{1}$ can be found simply as
\begin{equation}
\tau_{i_{1}i_{2} \ldots i_{N}}^{1} =\frac{1}{2} \left( \tau_{i_{1}i_{2} \ldots i_{N}}+\tau_{i_{1}i_{2} \ldots i_{N}}^{U_{1}}\right).
\end{equation}
We see therefore that $\tau_{00 \ldots}^{1} = \tau_{00 \ldots}$, $\tau_{33 \ldots}^{1} = \tau_{33 \ldots}$, $\tau_{03 \ldots}^{1} = 0$ and $\tau_{30 \ldots}^{1} = 0$.

Now, convexity and contractivity of the distance $D$ can be used to establish the inequality $D(\rho,\delta^{1}) \leq D(\rho,\delta)$ for any \NBD state $\rho$ and any incoherent state $\delta$. Indeed,
\begin{eqnarray}
D(\rho,\delta^{1}) &=& D\left(\rho,\frac{1}{2}(\delta+U_{1} \delta U_{1}^{\dagger})\right)  \\
&\leq& \frac{1}{2} \left( D(\rho,\delta)+D(\rho,U_{1} \delta U_{1}^{\dagger}) \right) \nonumber \\
&=& \frac{1}{2} \left( D(\rho,\delta)+D(U_{1} \rho U_{1}^{\dagger},U_{1} \delta U_{1}^{\dagger}) \right) \nonumber \\
&=& D(\rho,\delta),\nonumber
\end{eqnarray}
where in the first equality we use the definition of $\delta^{1}$, in the subsequent inequality we use the convexity of $D$, in the equality on the third line we use the invariance of $\rho$ through $U_{1}$, i.e. $U_{1} \rho U_{1}^{\dagger}=\rho$, and in the final equality we use the invariance of $D$ through unitaries $D(U_{1} \rho U_{1}^{\dagger},U_{1} \delta U_{1}^{\dagger})=D(\rho,\delta)$ implied by the contractivity of $D$.

Returning to the action of $U_{j}$ on $\delta$, it is a simple extension of the previous argument for $U_{1}$ to see that $U_{j}$ flips the value of $\tau_{i_{1}i_{2} \ldots i_{N}}$ when $i_{j} \neq i_{j+1}$. We are now in a position to define a set of incoherent states $\{\delta^{0},\delta^{1},\delta^{2} \ldots \delta^{N-1}\}$ in an iterative way
\begin{equation}
\delta^{j} = \frac{1}{2}\left( \delta^{j-1}+U_{j}\delta^{j-1}U_{j}^{\dagger}\right),
\end{equation}
for $j \in [1,N-1]$ and $\delta^{0} \equiv \delta$. The initial state is the incoherent state $\delta$ with correlation tensor elements $\tau_{i_{1}i_{2} \ldots i_{N}}$. The first state $\delta^{1}$ loses all the $\tau_{i_{1}i_{2} \ldots i_{N}}$ from $\delta$ for which $i_{1}\neq i_{2}$. Next, the $j$-th state $\delta^{j}$ loses all the $\tau_{i_{1}i_{2} \ldots i_{N}}$ from $\delta$ for which $i_{j}\neq i_{j+1}$. The final state $\delta^{N-1}$ loses all the $\tau_{i_{1}i_{2} \ldots i_{N}}$ from $\delta$ for which $i_{N-1} \neq i_{N}$. The only remaining values of $\tau_{i_{1}i_{2} \ldots i_{N}}$ from $\delta$ in $\delta^{N-1}$ are those for which $i_{1}=i_{2}=i_{3} \ldots = i_{N}$. Only $\tau_{00 \ldots 0}$ and $\tau_{33 \ldots 3}$ obey this condition, i.e. $\delta^{N-1}$ is the incoherent \NBD state $\delta_{{\rm{M}}^3_N}$.

The inequality $D(\rho,\delta^{1}) \leq D(\rho,\delta)$ can now be generalised iteratively,
\begin{eqnarray}
D(\rho,\delta^{j}) &=& D\left(\rho,\frac{1}{2}(\delta^{j-1}+U_{j} \delta^{j-1} U_{j}^{\dagger})\right)  \\
&\leq& \frac{1}{2} \left( D(\rho,\delta^{j-1})+D(\rho,U_{j} \delta^{j-1} U_{j}^{\dagger}) \right) \nonumber \\
&=& \frac{1}{2} \left( D(\rho,\delta^{j-1})+D(U_{j} \rho U_{j}^{\dagger},U_{j} \delta^{j-1} U_{j}^{\dagger}) \right) \nonumber \\
&=& D(\rho,\delta^{j-1}),\nonumber
\end{eqnarray}
where we use, in order, the definition of $\delta^{j}$ for $j \in [1,N-1]$, the convexity of $D$, the invariance of $\rho$ through any $U_{j}$, i.e. $U_{j} \rho U_{j}^{\dagger}=\rho$, and the invariance of $D$ through unitaries, $D(U_{j} \rho U_{j}^{\dagger},U_{j} \delta^{j-1} U_{j}^{\dagger})=D(\rho,\delta^{j-1})$.

This process gives a hierarchy of $N-1$ inequalities $D(\rho,\delta^{j}) \leq D(\rho,\delta^{j-1})$, which chained together imply $D(\rho,\delta^{0}) \leq D(\rho,\delta^{N-1})$. We know that $\delta^{0} \equiv \delta$ and $\delta^{N-1} \equiv \delta_{{\rm{M}}^3_N}$, hence we have shown that
\begin{equation}
D(\rho,\delta_{{\rm{M}}^3_N}) \leq D(\rho,\delta)\ \ \  \forall \delta.
\end{equation}
\begin{flushright}
$\blacksquare$
\end{flushright}

\begin{lemmaA}\label{lem:nearestclassicalBDstateinsidethexandzaxis}
For all even $N$, according to any contractive distance $D$, it holds that one of the closest incoherent \NBD states $\delta$ with triple $\{0,0,s\}$ to a \NBD state $\rho$ with triple $\{c_1,(-1)^{N/2}c_1c_3,c_3\}$ is specified by $s=c_3$.\label{eq:nearestNBDbelongingtothezaxis}
\end{lemmaA}
\noindent \textit{Proof}. We need to prove that, for any $z$, it holds that
\begin{eqnarray}
&&D(\{c_1,(-1)^{N/2}c_1 c_3,c_3\},\{0,0,c_3\}) \\ \nonumber
&\leq& D(\{c_1,(-1)^{N/2}c_1 c_3,c_3\},\{0,0,c_3+z\}).
\end{eqnarray}
In fact
\begin{eqnarray} \nonumber
& &D(\{c_1,(-1)^{N/2}c_1 c_3,c_3\},\{0,0,c_3\})  \\ \nonumber
&=& D(\{c_1,0,0\},\{0,0,0\})  \\ \nonumber
&=& D(\Lambda^{F_1 \otimes N}_{1}\{c_1,(-1)^{N/2}c_1 c_3,c_3\},\Lambda^{F_1 \otimes N}_{1}\{0,0,c_3+z\})  \\ \nonumber
&\leq& D(\{c_1,(-1)^{N/2}c_1 c_3,c_3\},\{0,0,c_3+z\}), \nonumber
\end{eqnarray}
where the first equality is due to Lemma A\ref{th:anycontractivedistanceismagictranslationalinvariant}, which holds for any contractive distance $D$ and any even $N$, the second equality is due to the fact that
\begin{eqnarray}
\{c_1,0,0\} &=& \Lambda^{F_1 \otimes N}_{1}\{c_1,(-1)^{N/2}c_1 c_3,c_3\}, \mbox{ and}\\
\{0,0,0\} &=& \Lambda^{F_1 \otimes N}_{1}\{0,0,c_3+z\},
\end{eqnarray}
with $\Lambda^{F_1 \otimes N}_{1}$ representing the action of $N$ local independent bit flip noisy channels expressed by Eq.~(\ref{eq:independentbitflipchannels}), when $k=1$ and $q=1$ (i.e., $t \rightarrow \infty$), and finally the inequality is due to contractivity of the distance $D$.
\begin{flushright}
$\blacksquare$
\end{flushright}

Due to Lemma A\ref{th:nearestincoherentstatetoaNBDstate} and Lemma A\ref{lem:nearestclassicalBDstateinsidethexandzaxis}, we finally get that any bona fide distance-based measure of quantum coherence ${\cal C}_{D}$ of the evolved \NBD state $\rho(q)$, given in Eq.~(\ref{eq:timeevolvedNBDstateunderlocalpuredephasing}), is equal to the following distance
\begin{eqnarray}\label{eq:frozenNqubitquantumcoherence}
{\cal C}_{D}(\rho(q))&=&D(\{c_1,(-1)^{N/2}(1-q)^N c_1c_3,(1-q)^N c_3\}, \nonumber \\
&&\quad \ \{0,0,(1-q)^N c_3\}),
\end{eqnarray}
which is frozen for any $q$ (equivalently, for any time $t$) thanks to Lemma A\ref{th:anycontractivedistanceismagictranslationalinvariant}, Eq.~(\ref{eq:magictranslationalinvariance2}).
 This concludes the proof of the central result in the main text.

 Notice further that Lemma A\ref{lem:nearestclassicalBDstateinsidethexandzaxis} implies that the $l_1$-norm of coherence equals (twice) the trace distance of coherence for all \NBD states with even $N$, which entails that ${\cal C}_{l_1}$ is frozen as well in the same dynamical conditions as for all bona fide distance-based measures of coherence, including e.g.~${\cal C}_{RE}$.



\begin{thebibliography}{81}
\bibitem[{not({\natexlab{a}})}]{notebasis}
\bibinfo{note}{For example, in a double slit experiment the basis is set by the
  chosen representation of the two possible paths; in quantum metrology
  \cite{Giovannetti2004}, the basis is set by the generator of the
  transformation that imprints the parameter onto the probe state; and in
  transport processes, including quantum biology \cite{Chin2012}, the reference
  basis is the eigenbasis of the Hamiltonian of the system.}
\bibitem[{not({\natexlab{b}})}]{notetime}
\bibinfo{note}{Specifically, for the dynamics of each qubit undergoing a flip
  channel, one has $q(t)=1-\exp(-\gamma t)$ where $t$ represents time and
  $\gamma$ is the decoherence rate \cite{Nielsen2010}.}
\bibitem[{not({\natexlab{c}})}]{noteflip}
\bibinfo{note}{We focus on bit flip channels $\Lambda^{F_1}$ and coherence in
  the computational basis (eigenbasis of $\sigma_3$) for illustration; however,
  analogous results hold for the other $k$-flip channels $\Lambda^{F_k}$ by
  considering coherence in the eigenbasis of $\sigma_{j \neq k}$.}
\end{thebibliography}

\end{document}